\documentclass[reviewcopy]{elsart}
\usepackage{graphics}
\usepackage{graphicx}
\usepackage{epsfig}
\usepackage{amssymb}

\begin{document}

\begin{frontmatter}


\title{Reaction mechanisms in the $^{6}$Li+$^{59}$Co system}
\author[USP]{F. A . Souza\corauthref{cor}\thanksref{TEL}},
\corauth[cor]{Corresponding author.}
\ead{fsouza@dfn.if.usp.br}
\author[IPHC]{C. Beck},
\author[USP]{N. Carlin},
\author[CEA]{N. Keeley\thanksref{ASI}}
\author[USP]{R. Liguori Neto}, 
\author[USP]{M. M. de Moura},
\author[USP]{M. G. Munhoz},
\author[USP]{M. G. Del Santo},
\author[USP]{A. A. P. Suaide},
\author[USP]{E. M. Szanto},
\author[USP]{A. Szanto de Toledo}
\address[USP]{Instituto de F\'{i}sica - Universidade de S\~ao Paulo, Departamento de F\'{i}sica Nuclear, C.P. 66318, 05315-970, S\~ao Paulo - SP, Brazil}
\address[IPHC]{Institut Pluridisciplinaire Hubert Curien, UMR 7178, CNRS-IN2P3 et Universit\'e Louis Pasteur, Bo\^{i}te Postale 28, F-67037 Strasbourg, Cedex 2, France}
\address[CEA]{CEA-Saclay DSM/IRFU/SPhN, F-91191 Gif sur Yvette Cedex, France}
\thanks[TEL] {phone: 55-11-3091-6939 fax: 55-11-3031-2742}
\thanks[ASI]{Permanent address: Department of Nuclear Reactions, The Andrzej So\l tan Institute for Nuclear Studies, ul.\ Ho\.za 69, PL-00681 Warsaw, Poland}

\begin{abstract}
The reactions induced by the weakly bound $^{6}$Li projectile interacting with the intermediate mass target $^{59}$Co were investigated. Light charged particles singles and $\alpha$-$d$ coincidence measurements were performed at the near barrier energies $E_{lab}=17.4, ~21.5, ~25.5$ and $29.6~{\rm MeV}$. The main contributions of the different competing mechanisms are discussed. A statistical model analysis, Continuum-Discretized Coupled-Channels calculations and two-body kinematics were used as tools to provide information to disentangle the main components of these mechanisms. A significant contribution of the direct breakup was observed through the difference between the experimental sequential breakup cross section and the CDCC prediction for the non-capture breakup cross section.
\end{abstract}
\end{frontmatter}

\section{Introduction}

Experiments with heavy ions performed during the last decade have shown that the internal degrees of freedom of the interacting nuclei play an important role in determining the reaction flux diverted toward the fusion reaction~\cite{1,2,3,4,5,5b}. Barrier distribution measurements~\cite{3} have shown that the coupling of collective degrees of freedom to the fusion channel may enhance the sub-barrier total fusion cross section. Interest in fusion studies at near- and sub-barrier energies with exotic nuclei as projectiles~\cite{5,5b,6,6b,7,8,9,9b,10,11} has been renewed with the recent increased availability of Radioactive Ion Beams (RIB).
The investigation of such reactions involving either unstable nuclei, far from the valley of
stability, or weakly bound stable nuclei, such as $^{6}$Li, should have a great impact on the study of astrophysical processes at very low bombarding energies near the Gamow peak~\cite{11,12}.
Light weakly bound stable and unstable nuclei display low nucleon (cluster) separation energies, and are therefore candidates for important breakup (BU) cross sections.

This possibility affects the dynamics of fusion reactions~\cite{13,14,15,16,17,18,19,20} due to the fact that part of the incoming flux may be lost from the entrance channel before overcoming the fusion barrier and, moreover, one of the fragments removed from the projectile (or target) may fuse leading to an important incomplete fusion (ICF) or transfer (TR) contribution. Following the review paper of Canto \textit{et al}.~\cite{4}, we consider here that ICF is a two-step process. After the breakup of the projectile one of the fragments, with approximately the projectile velocity, interacts with the target leading to a compound system formation. On the other hand, TR would be a one-step process in which there is a transfer of a fragment from the projectile to unbound states of the target followed by a particle evaporation. The final residual nucleus is the same in both cases, being a challenge for the experimental separation of these processes.

The contributions of these reaction mechanisms have not so far been identified in barrier distribution measurements or clearly disentangled in ``singles'' (inclusive) particle measurements. 

Coincidence (exclusive) measurements are required to guarantee the occurrence of BU processes in order to shed some light on the understanding of this problem which remains controversial, as conflicting theoretical expectations have been reported in the recent past~\cite{21,22,23,24,25,Diaz03,27,28,29}.

We have already performed measurements for $^{6,7}$Li beams incident on the intermediate-mass target $^{59}$Co at near barrier energies and studied the total fusion~\cite{Beck03}, elastic scattering~\cite{31} and BU cross sections~\cite{32}. In this work we present a study of both inclusive and exclusive light charged particle (LCP) energy spectra for the $^{6}$Li~+~$^{59}$Co system and the respective contributions of the different mechanisms are discussed. Measurements were performed at four bombarding energies above the Coulomb barrier ($V_{B} = 12.0$ MeV). Experimental details are given in Sec.\ 2. A statistical-model analysis and two-body kinematics, presented in Sec.\ 3, were used as tools of an attempt to distinguish complete fusion (CF), ICF, TR and BU components and to provide information on their respective properties. Sec.\ 3 proposes a discussion of the cross section balance assuming that the BU yield can be estimated within the Continuum-Discretized Coupled-Channels (CDCC) approach~\cite{Diaz03,27,28,29}. Also in this section we discuss the sequential breakup cross section for the first excited state $3^{+}$ ($E^{*}=2.186$~MeV) of $^{6}$Li obtained from the $\alpha$-$d$ coincidence analysis.

\section{Experimental details}

The experiments were performed at the University of S\~{a}o Paulo Physics Institute. The $^{6}$Li beam was delivered by the 8UD Pelletron accelerator with energies $E_{lab}=18, ~22, ~26$ and $30$~MeV, and bombarded a $2.2$~mg/cm$^{2}$ thick $^{59}$Co target. Due to the target thickness the bombarding energies were corrected for the energy loss at the center of the target. The corrected energies are $E_{lab} = 17.4,~21.5,~25.5$ and $29.6$~MeV, respectively.

The LCPs emitted during the $^{6}$Li~+~$^{59}$Co reaction were detected by means of 11 triple telescopes~\cite{33} separated by $\Delta \theta=10^{\circ}$ and installed in the reaction plane. The triple telescopes were composed of an ionization chamber followed by a $150~\mu$m Si(SB) detector and a $40$~mm CsI crystal with photodiode readout to measure the LCP residual energy. The entrance window of the ionization chamber was a $150~\mu$g/cm$^{2}$ aluminized polypropylene film. The use of $20$~torr isobutane in the ionization chambers allowed an energy resolution of $7.6\%$ in their respective signals.

\begin{figure}
\begin{center}
\includegraphics[width=1.0\columnwidth]{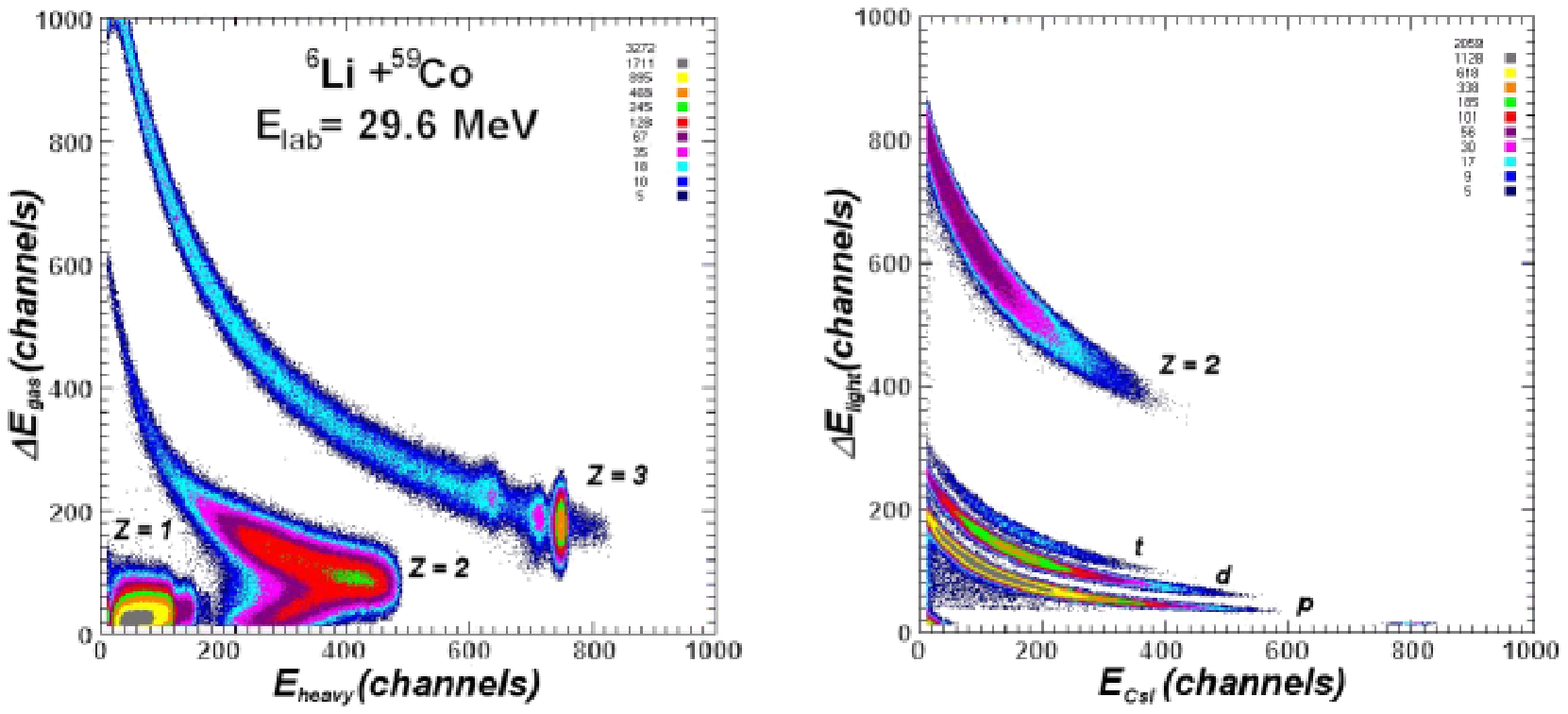}
\caption{Two-dimensional spectra of the $\Delta E_{gas} \times E_{heavy}$ and $\Delta E_{light} \times E_{CsI}$ for $^{6}$Li+$^{59}$Co at $E_{lab}=29.6$ MeV. In the $\Delta E_{light} \times E_{CsI}$ spectrum note the clear separation of the different isotopes for $Z=1$.}
\label{fig:biparametric}
\end{center}
\end{figure}

Identification of the LCPs emitted during the reaction was achieved by means of two-dimensional spectra of the $\Delta E_{gas}$, $E_{heavy}$, $\Delta E_{light}$ and $E_{CsI}$ signals (see Fig.~\ref{fig:biparametric}) processed by means of standard NIM and CAMAC electronics. The $\Delta E_{light} \times E_{CsI}$ spectrum in Fig.~\ref{fig:biparametric} clearly shows the high-quality of the mass-discrimination for H isotopes ($p,d,t$). The $\Delta E_{gas}$ comes from the ionization chamber. The $E_{heavy}$ and $\Delta E_{light}$ signals are generated by the Si detector with low- and high-gain, respectively, and the $E_{CsI}$ signal represents the residual energy deposited in the CsI crystal. The energy loss in each detector was calculated using a universal analytic equation~\cite{34}. The $\Delta E_{gas}$ and $E_{heavy}$ signals were calibrated using the $^{6}$Li elastic scattering peaks. The curves of the residual energy deposited in the CsI crystal as a function of energy loss in the Si detector for each $Z$ and the linear relation between the $E_{heavy}$ and $\Delta E_{light}$ gains were used to calibrate the energy spectra of the LCPs. The telescopes covered the angular range from $\theta = -45^{\circ}$ to $\theta = -15^{\circ}$ and from $\theta = 15^{\circ}$ to $\theta = 75^{\circ}$, both in $\Delta \theta = 10^{\circ}$ steps. The solid angles of the telescopes varied from $\Delta \Omega = 0.14$ to $\Delta \Omega = 1.96$~msr. Absolute cross sections were determined from our earlier elastic scattering measurements~\cite{31}.

Some details of part of this experimental setup description can also be found in Refs.~\cite{31,33}.

\section{Results and discussion}

For reactions induced by the weakly bound projectile $^{6}$Li ($Q = -1.47$~MeV for the $\alpha+d$ breakup) it is natural to assume that the main contributor to the $\alpha$ and $d$ yields is the $\alpha+d$ breakup, but other processes are also likely to occur with significant cross sections~\cite{20}. The processes we take into account are the following:\\

a) $^{6}$Li + $^{59}$Co $\rightarrow$ $^{6}$Li$^{*}$ + $^{59}$Co
$\rightarrow$  $\alpha+d+~^{59}$Co

b) $^{6}$Li + $^{59}$Co $\rightarrow$ $\alpha$ + $^{61}$Ni$^{*}$
$\rightarrow$ subsequent decay

c) $^{6}$Li + $^{59}$Co $\rightarrow$ $d$ + $^{63}$Cu$^{*}$
$\rightarrow$ subsequent decay

d) $^{6}$Li + $^{59}$Co $\rightarrow$ $^5$Li + $^{60}$Co$^{*}$
$\rightarrow$ subsequent decay

e) $^{6}$Li + $^{59}$Co $\rightarrow$ $^5$He + $^{60}$Ni$^{*}$
$\rightarrow$ subsequent decay

f) $^{6}$Li + $^{59}$Co $\rightarrow$ $^{65}$Zn$^{*}$
$\rightarrow$ subsequent decay \\

Process a) is identified as the breakup of $^{6}$Li, which could be either direct or resonant (sequential). In this case there is no further capture of the BU products by the target; following the definitions of Ref.~\cite{4}, we will call it non-capture breakup (NCBU). Process b) is identified as either ICF of $d$+$^{59}$Co ($d$-ICF) after BU or a direct one-step $d$ transfer ($d$-TR), both with subsequent decay of the excited $^{61}$Ni$^{*}$. Here, the $\alpha$ particle is left as a ``spectator''. In the same way, process c) can be identified as either ICF of $\alpha$+$^{59}$Co ($\alpha$-ICF) after BU or a direct one-step $\alpha$ transfer ($\alpha$-TR), both with subsequent decay of the excited $^{63}$Cu$^{*}$. In this case the $d$ is left as a ``spectator''. Processes d) and e) represent single neutron and single proton stripping from the $^6$Li projectile, respectively with subsequent decay of the unstable $^5$Li and $^5$He leaving an $\alpha$ particle plus a neutron or proton. Process f) is simply identified as complete fusion (CF). In all processes involving deuteron emission in the exit channel subsequent breakup of the deuteron was not taken into account, in accordance with Refs.~\cite{45,Pakou07}.

Our experimental setup allowed us to obtain both ``singles'' LCP and coincidence LCP data. First, we will concentrate on the results obtained from the analysis of the ``singles'' LCP data and, finally we discuss the analysis of the $\alpha$-$d$ coincidence data which was used to obtain the sequential breakup cross section.

\begin{figure}
\begin{center}
\includegraphics[width=1.0\columnwidth]{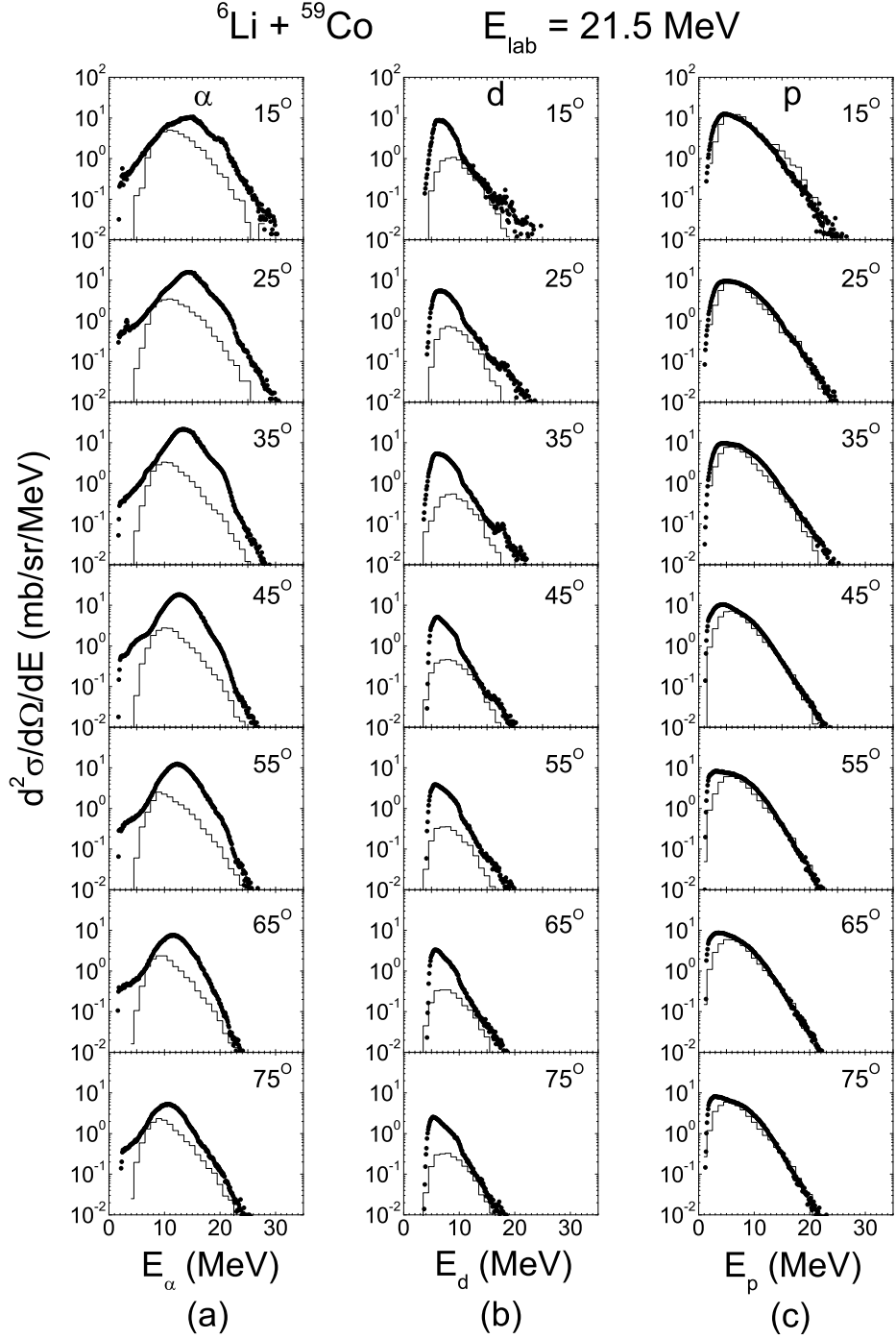}
\caption{(a) Experimental inclusive $\alpha$ energy spectra for $E_{lab} = 21.5$~MeV, at $\theta_{lab}=$ 15, 25, 35, 45, 55, 65 and 75 degrees, and the respective CACARIZO predictions (histograms) for the CF decay. (b) and (c) the same for $d$ and $p$, respectively. The error bars are of the same size or smaller than the symbols used to represent the experimental points.}
\label{fig:spectra}
\end{center}
\end{figure}

In Fig.~\ref{fig:spectra} we show singles $\alpha$, $d$ and $p$ production spectra for $E_{lab}=21.5$~MeV and at $\theta_{lab}=$ 15, 25, 35, 45, 55, 65 and 75 degrees (columns a, b and c respectively) together with statistical-model predictions for CF decay (histograms) using the Hauser-Feshbach evaporation code CACARIZO~\cite{35,35b} (the Monte Carlo version of CASCADE~\cite{35b}). In the calculations the transmission coefficients were evaluated using optical model (OM) parameters for spherical nuclei. The compound nucleus (CN) angular momentum distributions were specified using the diffuseness parameter $\Delta L=1$ and the critical angular momentum $L_{crit}$ calculated internally by the code for each bombarding energy. The OM potentials for $n$, $p$, and $\alpha$ were taken from Rapaport \textit{et al}.~\cite{36}, Perey~\cite{37}, and Huizenga and Igo~\cite{38}, respectively. One of the most important parameters in the calculations is the level density parameter $a$. In our case it was defined as $a_{LDM} = A/10$~\cite{39} rather than the A/8 value adopted for other systematic studies~\cite{35b}. This value of $a$, needed to reproduce the Giant Dipole Resonance (GDR) enhancement in the $^{6}$Li~+~$^{57}$Fe $\gamma$-ray spectra~\cite{39}, provided good results for the LCP energy spectra without any extra normalization on the CF cross sections. In particular, the proton energy spectra for which we expect essentially CN decay (except in the low-energy region where $p$ decay from ICF and TR intermediate nuclei might be apparent; protons from $d$ breakup were not considered, as already argued) were well reproduced for all detection angles (as shown in Fig.~\ref{fig:spectra}c). We performed additional CACARIZO calculations for $d$- and $\alpha$-ICF assuming bombarding energies corresponding to the $^{6}$Li projectile velocity. The location of the $p$ decay energies supports well this rather crude hypothesis.

\begin{figure}
\begin{center}
\includegraphics[width=1.0\columnwidth]{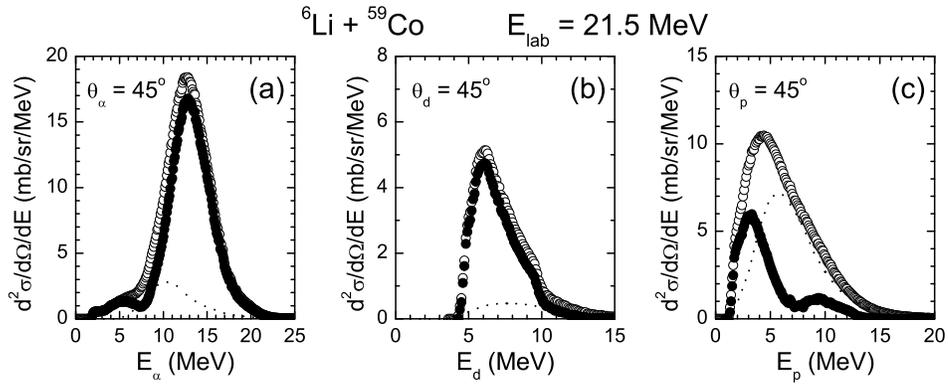}
\caption{(a) Experimental inclusive $\alpha$ energy spectrum (open circles) and $\alpha$-bump (full circles) at $\theta=45^{\circ}$ for $E_{lab} = 21.5$~MeV,
obtained after subtracting the contribution of CF decay, as calculated by CACARIZO (dotted line). (b) and (c) The same for $d$ and $p$ respectively.}
\label{fig:linearspectra}
\end{center}
\end{figure}

Fig.~\ref{fig:linearspectra} displays energy spectra at $\theta$ = $45^{\circ}$ for $E_{lab} = 21.5$~MeV, using a linear scale in the $y$-axis. The same energy spectra are given in a log scale in Fig.~\ref{fig:spectra} for all the possible detection angles. Very similar spectra (not shown) were obtained for the other bombarding energies at different angles. In this figure we note that there is a contribution from other mechanisms in the LCP production spectra (open circles). For $p$ (Fig.~\ref{fig:linearspectra}c), after the subtraction of the evaporative component of CF (dotted line obtained from CACARIZO) the major contributions remaining (full circles) at lower energies may be attributed mainly to decay of ICF and TR intermediate nuclei. 
One should also take note that deformation effects and lowering of $p$ emission barriers~\cite{35b}, not explicitly taken into account in the present CACARIZO calculations, might also explain the large yields observed at low energies. 
The high energy $p$ can not be attributed to ICF or TR and also these energies are not related to the projectile velocity. As discussed in Ref.~\cite{Badran01} it may correspond to some sort of pre-equilibrium process. For $\alpha$ particles, after subtraction of the contribution from the CF $\alpha$ particles as calculated by CACARIZO, two ``bumps'' remain, as can be seen in Fig.~\ref{fig:linearspectra}a. In the same figure the small low-energy bump is attributed to decay of ICF and TR intermediate nuclei. This attribution is supported by the results of the CACARIZO calculations for $d$ and $\alpha$-ICF. The high-energy bump is the subject of the analysis that follows.

For the high-energy $\alpha$-bump, according to the previous description, we are then dealing with the experimental quantity $\sigma_{\alpha-bump}$ defined as:

\begin{equation}
\sigma_{\alpha-bump} = \sigma_{d-ICF} + \sigma_{d-TR} + \sigma_{NCBU} + \sigma_{n-TR} + \sigma_{p-TR}
\end{equation}

Analogously for the $d$ singles energy spectra, shown in Fig.~\ref{fig:linearspectra}b, we may define the quantity $\sigma_{d-bump}$ as:

\begin{equation}
\sigma_{d-bump} = \sigma_{\alpha-ICF} + \sigma_{\alpha-TR} + \sigma_{NCBU}
\end{equation}

\begin{figure}
\begin{center}
\includegraphics[width=1.0\columnwidth]{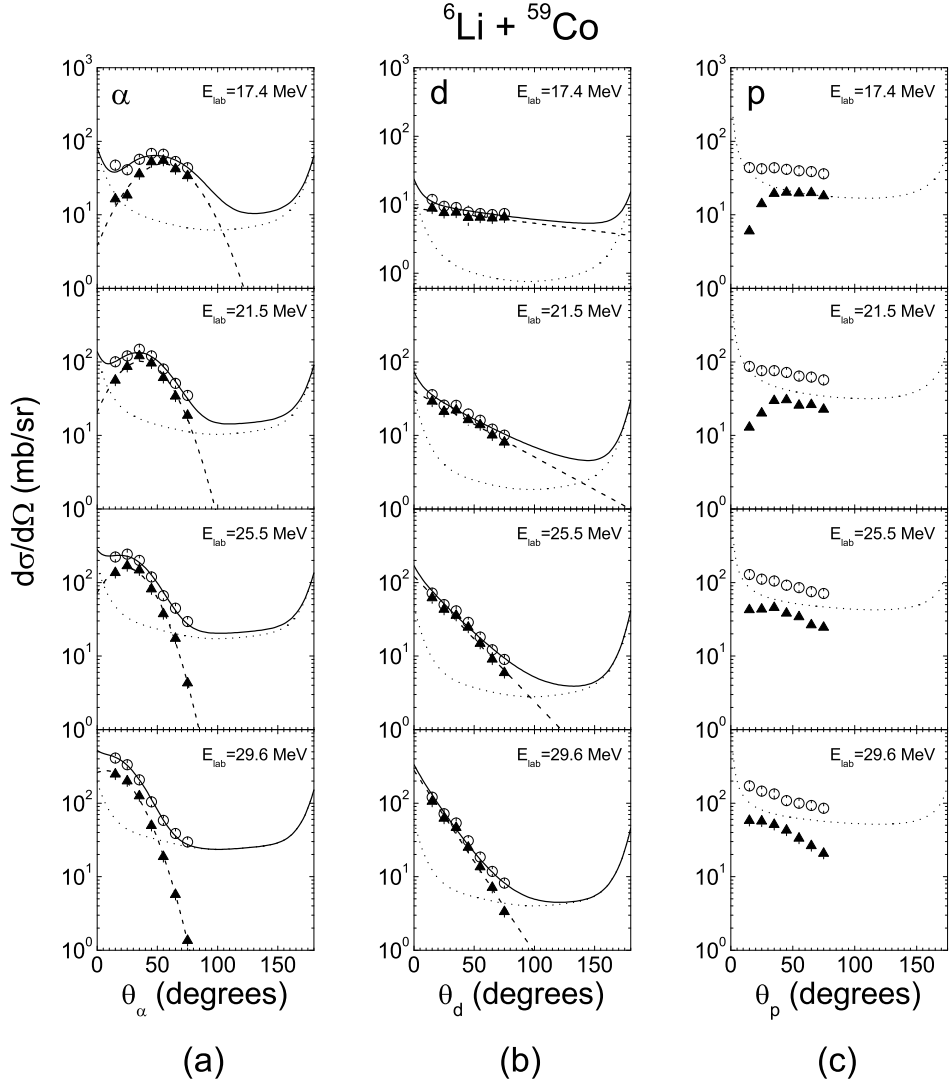}
\caption{(a) (From top to bottom) Angular distributions for the total $\alpha$ production (open circles) and high-energy $\alpha$-bump (full triangles) for $E_{lab} = 17.4,~21.5,~25.5,~29.6$~MeV. (b) and (c) The same for $d$ and $p$. The solid and dashed lines correspond to the shapes adopted for the integration of the angular distributions. The dotted line is the CACARIZO prediction for CF decay. In most cases the error bars are of the same size or smaller than the symbols used to represent the experimental points. For $p$, the full triangles represent the difference between total production and CF decay predictions.}
\label{fig:angdist}
\end{center}
\end{figure}

The quantities $\sigma_{\alpha-bump}$ and $\sigma_{d-bump}$ were obtained through the integration of the angular distributions (dashed lines) shown in Fig.~\ref{fig:angdist}a and Fig.~\ref{fig:angdist}b, respectively, for all bombarding energies. The dashed lines were obtained by means of Gaussian fits to the experimental data in the case of $\alpha$-bump and exponential fits for $d$-bump. In the same figures we present experimental $\alpha$, $d$ and $p$ angular distributions. As we only have data points up to $\theta$ = $75^{\circ}$ we have assumed that the total $\alpha$ and $d$ production at backward angles is essentially due to CF and ICF/TR decays. In order to estimate the shape of the angular distribution for the backward angles we used CACARIZO predictions for the CF decay. The adopted shapes are consistent with published data for $^{6}$Li~+~$^{58}$Ni at similar bombarding energies~\cite{Pfeiffer73}. As explained earlier, due to the non CF decay contributions, the angular distributions for $p$ are not reproduced by CACARIZO predictions (Fig.~\ref{fig:angdist}c), which is also consistent with the discussion about the ICF/TR decay adopted in this work.

\begin{figure}
\begin{center}
\includegraphics[width=1.0\columnwidth]{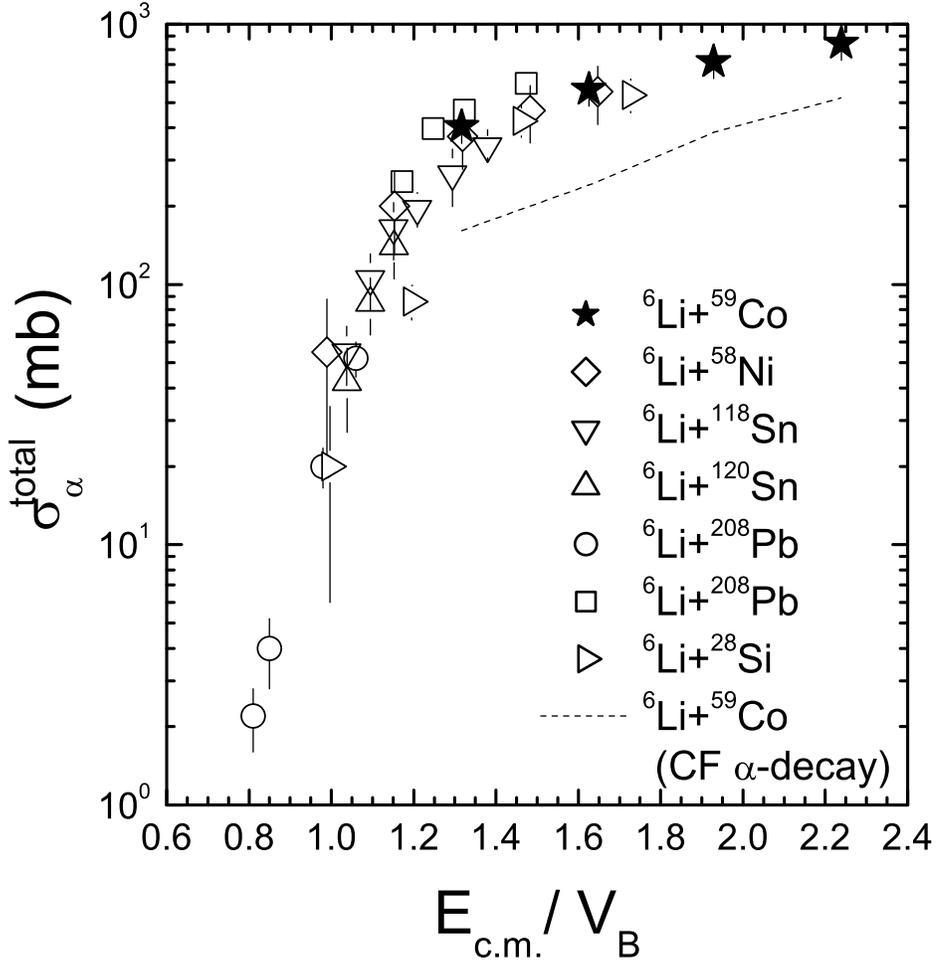}
\caption{Total $\alpha$ production cross sections in reactions involving $^{6}$Li on various targets as a function of the center of mass energy divided 
by the Coulomb barrier energy. We incorporate results extracted from~\cite{Pfeiffer73,Pakou03} and from~\cite{20} for the $^{208}$Pb target (open squares). We also include our results for 
$^{6}$Li~+~$^{59}$Co, which reproduce well the universal behavior of $\alpha$ production. The dashed line indicates the cross sections for $\alpha$ 
particles evaporated during the $^{6}$Li~+~$^{59}$Co CF process as simulated by the CACARIZO evaporation code.}
\label{fig:totalalpha}
\end{center}
\end{figure}

In Fig.~\ref{fig:totalalpha} we present an excitation function, adopted from~\cite{Pfeiffer73,Pakou03}, of total $\alpha$ production cross section as a function of reduced energy for $^{6}$Li on various targets at near and above barrier energies~\cite{20,Pfeiffer73,Pakou03}. As noted in Ref.~\cite{Pakou03}, a simple  systematic behavior for total $\alpha$ production is observed with no significant target dependence. We also include the present results for $^{6}$Li~+~$^{59}$Co, obtained from the integration of the angular distributions (i.e.\ the solid curve in Fig.~\ref{fig:angdist}a). The Coulomb barrier ($V_{B}=12.0$ MeV) was extracted from Ref.~\cite{Beck03}. We note that the $^{6}$Li~+~$^{59}$Co data also obey the systematic trend giving further support to the present analysis. It is worth noting that a similar trend has been obtained for $^{7}$Li projectiles~\cite{Pakouli7}. 
For the sake of comparison, we have plotted in Fig.~\ref{fig:totalalpha} (dashed line) the excitation function of $\alpha$ particles calculated by CACARIZO for $^{6}$Li~+~$^{59}$Co reaction, i.e. all the $\alpha$ particles that are emitted through a CF evaporation process.
As the experimental data (stars) lie well above the fusion predictions we may conclude that the ICF and TR components both play a significant role in the total $\alpha$ production. This behavior is even stronger for $^{6}$He induced reactions~\cite{6,7,8,40} for which the measured total $\alpha$ cross sections are much larger than for $^{6}$Li due to the strong competition of the 1n- and 2n-transfer reactions as convincingly demonstrated in the $^{6}$He~+~$^{209}$Bi system~\cite{41,41b}, for instance.

A clear separation of mechanisms involves a knowledge of the $\sigma_{NCBU}$ cross section. The NCBU is the sum of the direct and sequential breakup processes. The non-model dependent analysis of the experimental data for direct breakup processes is a very difficult task and work is in progress to accomplish such a challenge~\cite{Thesis}. Thus, in this work we adopted the approach of performing CDCC~\cite{Diaz03,27,28,29,44} calculations to evaluate $\sigma_{NCBU}$ and sequential breakup cross sections for the first excited state $3^{+}$ of $^{6}$Li ($\sigma_{3^{+}}$). The exclusive BU cross sections for the resonant states in $^{6}$Li plus the non-resonant $\alpha$+$d$ continuum were calculated using a cluster-folding model with potentials that describe well the measured elastic scattering angular distributions~\cite{27,28,29}. The CDCC calculations for $^{6}$Li were performed with the code FRESCO assuming an $\alpha+d$ cluster structure, similar to that described in Refs.~\cite{Diaz03,27}. The $\alpha+d$ binding potentials were taken from~\cite{42} and couplings to the $3^{+}$ ($E^{*}=2.18$~MeV), $2^{+}$ ($E^{*}=4.31$~MeV) and $1^{+}$ ($E^{*}=5.65$~MeV) resonant states were included as well as couplings to the non-resonant $\alpha+d$ continuum. The continuum was discretized into a series of momentum bins of width $\delta k = 0.2$~fm$^{-1}$ with maximum $k=1$~fm$^{-1}$, where $\hbar k$ denotes the momentum of the $\alpha+d$ relative motion. In order to avoid double counting the width $\delta k$ was suitably modified in the presence of resonances. In the calculations each momentum bin was treated as an excited state of $^{6}$Li, at an excitation energy equal to the mean energy of the bin and having spin $\vec{I}$ and parity $(-1)^{L}$. The angular momenta are related by $\vec{I}=\vec{L}+\vec{s}$, where $\vec{s}$ is the spin of the $d$ and $\vec{L}$ is the relative angular momentum of $\alpha+d$ cluster system. Following Hirabayashi~\cite{43} couplings to states with $L\geq3$ are expected to be small. Thus, $L$ was limited to $0, ~1, ~2, ~3$. All couplings, including continuum-continuum couplings, up to multipolarity $\lambda=3$ were included. Details of the CDCC method may be found in Refs.~\cite{Diaz03,27,28,29,44}.

In Table~\ref{tab:xsections} we present a summary of our results obtained from
the experimental LCP singles spectra and the evaluation of non-capture BU (NCBU)
cross sections with CDCC~\cite{27}. The total reaction cross sections were
extracted from our elastic scattering analysis~\cite{31} using the S\~ao Paulo
Potential~\cite{44a} and from the CDCC calculations~\cite{27}. The OM fits and
the CDCC calculations yield similar cross sections which are much larger than
the total fusion cross sections~\cite{Beck03} measured at $E_{lab} = 17.4$~MeV
and $E_{lab} = 25.5$~MeV using the gamma-ray method~\cite{Beck03}. Let us recall
that the measured total fusion cross sections were also found to be rather
well reproduced by the CDCC method~\cite{Diaz03,44}. However, some unexpected 
discrepancy can be observed in Table I for the lowest energy. Although this
problem may appear to be still open, one may propose two possible explanations:
            i) due to the limitations of the gamma-ray method, the
            experimental total fusion cross section might have been
            underestimated in Ref.\cite{Beck03}
            ii) cross sections values as predicted by CDCC in
            Ref.\cite{27} are, somehow, quite large.

When comparing the values of $\sigma_{\alpha-bump}$ and $\sigma_{d-bump}$ in Table~\ref{tab:xsections} we note that there is an excess of $\alpha$ particles over $d$ (approximately a factor of $3$). In the case of $^{6}$Li~+~$^{28}$Si reaction a very good qualitative agreement has been found for the large TR cross sections as compared with DWBA calculations~\cite{Pakou07}. This behavior for a $^{59}$Co target confirms that found previously for $^{58}$Ni and $^{118,120}$Sn targets~\cite{Pfeiffer73} at similar bombarding energies. 
Single nucleon transfer reactions will also produce $\alpha$ particles but not deuterons, and thus could also contribute to the excess of $\alpha$ particles over deuterons. Although a full calculation of these processes is not possible for a $^{59}$Co target due to the high density of states in the residual target-like nuclei, DWBA estimates suggest that the single nucleon transfer cross sections are at least as large as those for NCBU~\cite{27}. A similar excess of $\alpha$ particles over $d$ has also been reported previously in the literature for other systems, not only for energies similar to ours~\cite{Pfeiffer73} but also at higher energies~\cite{45,46}.

The results presented in Table~\ref{tab:xsections} (note that the CDCC cross sections reported in Table~\ref{tab:xsections} were obtained by interpolation of the values calculated at 18, 26, and 30 MeV in Ref.~\cite{27}) show that the NCBU cross section is significantly lower than the $\sigma_{\alpha-bump}$ and $\sigma_{d-bump}$ cross sections. This is also observed in another work~\cite{27}. In this case we could argue that the main contributions to $\sigma_{\alpha-bump}$ and $\sigma_{d-bump}$ are most probably due to both the ICF and TR mechanisms.

\begin{figure}
\begin{center}
\includegraphics[width=1.0\columnwidth]{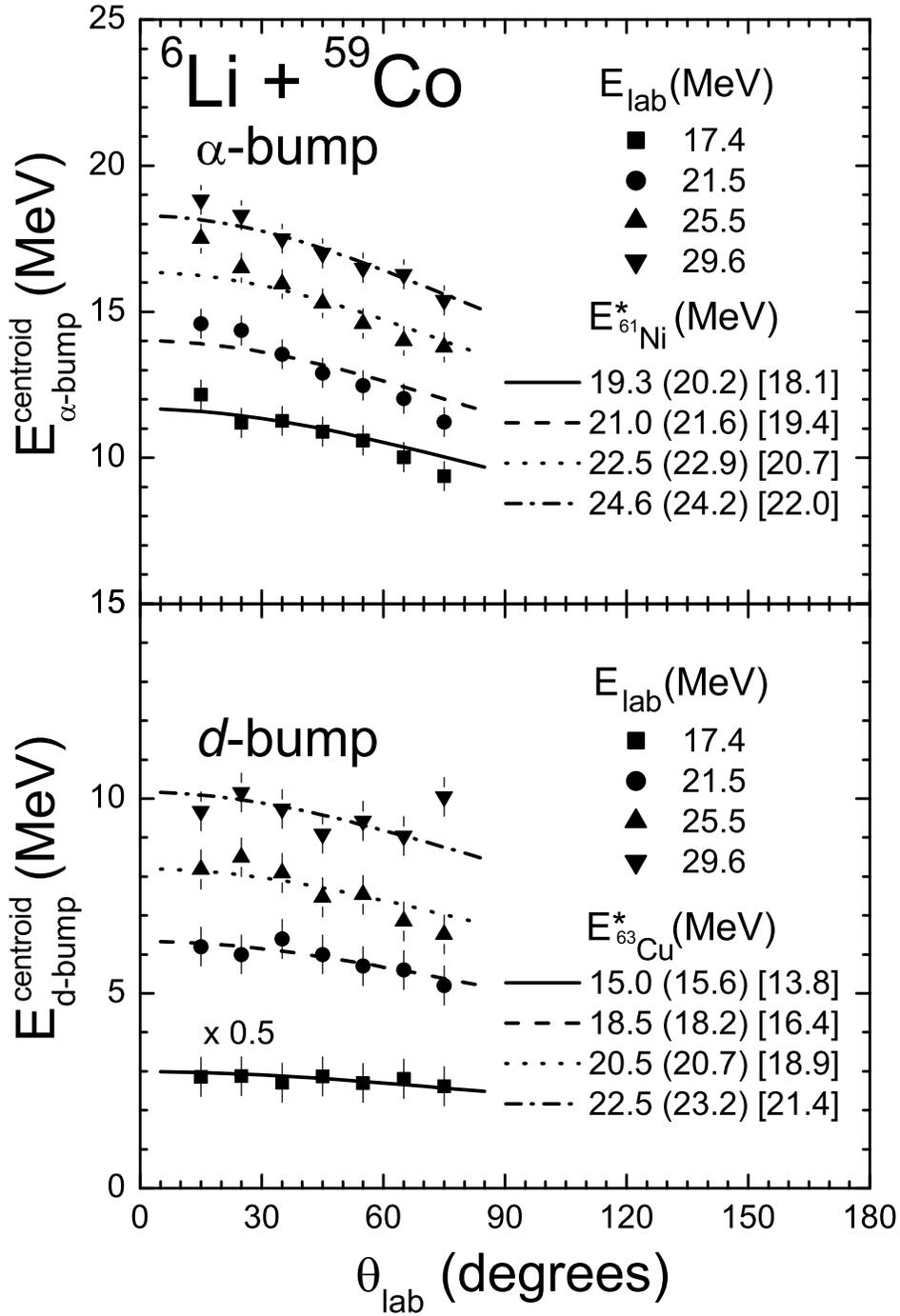}
\caption{Energy of the centroids of the $\alpha$-bump and $d$-bump as a function of the detection angle for all bombarding energies. The curves
are two-body kinematics results and represent the behavior of the excited nuclei that provided the best fits to 
the experimental results. The values between parentheses are the calculated excitation energies for the exit-channel nuclei formed in an ICF process. 
The values between brackets are the calculated excitation energies in the exit fragments as formed in a direct TR process.}
\label{fig:kinematics}
\end{center}
\end{figure}

In order to confirm whether our assumption is reasonable we performed a two-body kinematics analysis related to the centroids of the high-energy $\alpha$-bump and $d$-bump as a function of the detection angle. For the sake of simplicity we have not considered three-body kinematics calculations which would have to be performed for the TR processes labeled d) and e). If the ICF and TR mechanisms are dominant the energy corresponding to the centroids should reflect the excitation energy of the $^{61}$Ni$^{*}$ and $^{63}$Cu$^{*}$ nuclei formed in the intermediate stage of processes b) and c) described above, as they are two-body processes. In Fig.~\ref{fig:kinematics} we show the behavior of the energy associated with the centroids of the high-energy $\alpha$-bump and the $d$-bump for all bombarding energies. We also present two-body kinematics calculations for the $\alpha$ and $d$ energies as a function of the detection angle for fixed excitation energies of $^{61}$Ni$^{*}$ and $^{63}$Cu$^{*}$. The uncertainty in the particle energy corresponds to the uncertainty in the determination of the total energy  ($\sim0.5$~MeV). The different curves in Fig.~\ref{fig:kinematics} represent the behavior of the excited nuclei that provided the best fits to the experimental results. The uncertainty associated with the fits is approximately $0.5$~MeV. The good agreement with the experimental results suggests that our assumption about the mechanisms is reasonable.

Considering the experimental uncertainties the excitation energies obtained are consistent with an ICF process for which the $\alpha$ and $d$ have approximately the projectile velocity. The calculated values are shown between parentheses in Fig.~\ref{fig:kinematics}. On the other hand, if we consider the TR process the agreement between the best experimental excitation energies and the ones obtained from optimum $Q$-value calculations~\cite{47} (shown between brackets in Fig.~\ref{fig:kinematics}) is not as good as for the ICF case. Due to the existence of different relations for calculating optimum $Q$-values we cannot a priori rule out the contribution of the TR processes labelled d) and e). The neutron TR contribution, for instance, has been found to be a rather competitive reaction channel in the $^{6}$Li~+~$^{118}$Sn and $^{6}$Li~+~$^{208}$Pb reactions~\cite{Scholz77} as well as in the $^{6}$Li~+~$^{28}$Si reaction~\cite{Pakou07}. It is worth noting that following Ref.~\cite{45,Pakou07} we did not consider the secondary disintegration of the deuterons, the contribution of which is expected to be much smaller~\cite{45}.

From this analysis we conclude that the main contributions to the $\alpha$-bump and $d$-bump are due to both ICF and TR. However, it was not possible to disentangle their individual contributions from the present inclusive data.

\begin{figure}
\begin{center}
\includegraphics[width=1.0\columnwidth]{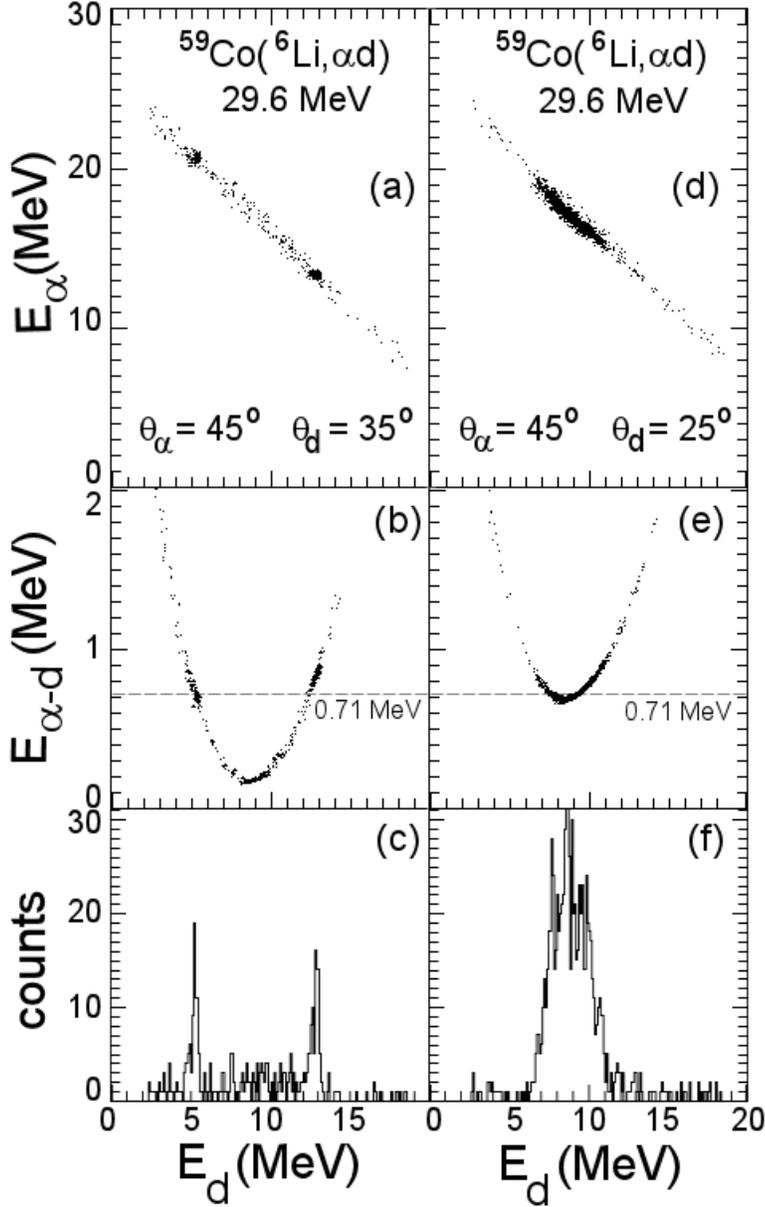}
\caption{Experimental $\alpha$-$d$ coincidence spectra for $^{6}$Li~+~$^{59}$Co at $E_{lab}=29.6$ MeV. The events were restricted to $Q=-1.475$ MeV. (a) $\alpha$-particle energy ($E_{\alpha}$) as a function of $d$ energy ($E_{d}$) for $\theta_{\alpha}=45^{\circ}$ and $\theta_{d}=35^{\circ}$. (b) The $\alpha$-$d$ relative energy ($E_{\alpha-d}$) as a function of $E_{d}$. For $E_{\alpha-d}=0.71$~MeV we can notice two peaks which correspond to the first excited state $3^{+}$ ($E^{*}=2.186$~MeV) of $^{6}$Li. (c) Projection of the events observed in (a) or (b) in the $E_{d}$ axis. (d), (e) and (f) The same for $\theta_{\alpha}=45^{\circ}$ and $\theta_{d}=25^{\circ}$.
}
\label{fig:spect_coinc}
\end{center}
\end{figure}

In the following part of this work, we will focus on the determination of sequential breakup cross section for the $3^{+}$ state of $^{6}$Li ($\sigma^{\rm exp}_{3^{+}}$). With the same experimental setup described previously, we have performed $\alpha-d$ coincidence measurements for each pair of detectors and took into account the events with $Q=-1.475$~MeV ($^{6}{\rm Li}\rightarrow\alpha+d$). Typical $\alpha-d$ coincidence spectra are shown in Fig.~\ref{fig:spect_coinc} for $^{6}$Li~+~$^{59}$Co at $E_{lab}=29.6$~MeV. The two peaks in the $E_{\alpha} \times E_{d}$ spectrum for $\theta_{\alpha}=45^{\circ}$ and $\theta_{d}=35^{\circ}$ (Fig.~\ref{fig:spect_coinc}a) correspond to the first excited state $3^{+}$ ($E^{*}=2.186$~MeV) of $^{6}$Li with a relative energy of $E_{\alpha-d}=0.71$~MeV as can be observed in Fig.~\ref{fig:spect_coinc}b. The projection of these events in the $E_{d}$ axis is shown in Fig.~\ref{fig:spect_coinc}c. Although the two peaks have the same $E_{\alpha-d}$ relative energy, they represent two different emission angles of the $^{6}$Li$^{*}$ decay. The same excited state of $^{6}$Li is observed in Fig.~\ref{fig:spect_coinc}d to \ref{fig:spect_coinc}f for $\theta_{\alpha}=45^{\circ}$ and $\theta_{d}=25^{\circ}$. In this case, the two peaks are very close due to kinematical limits of the breakup detection cone. Please note that similar coincidence spectra have been measured for the three other bombarding energies.

\begin{figure}
\begin{center}
\includegraphics[width=1.0\columnwidth]{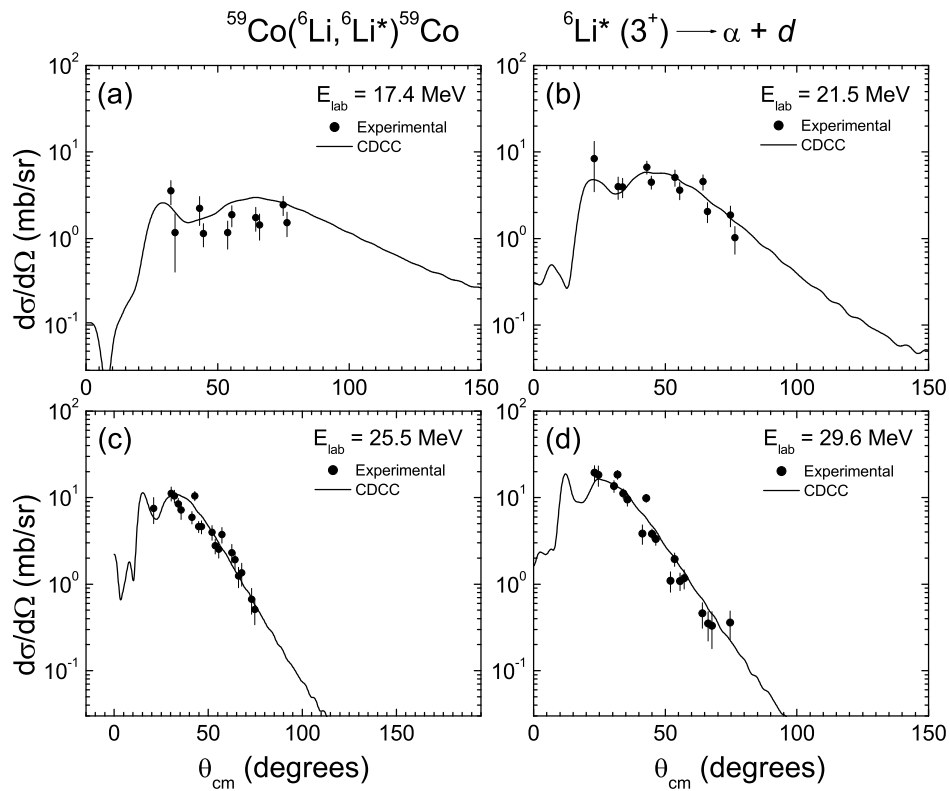}
\caption{Experimental angular distributions for the first excited state $3^{+}$ ($E^{*}=2.186$~MeV) of $^{6}$Li (full circles) and corresponding CDCC calculations (solid
lines).}
\label{fig:BUdistrib}
\end{center}
\end{figure}

The experimental sequential breakup angular distributions (full circles in Fig.~\ref{fig:BUdistrib}) were determined considering the $d$ and $\alpha$-particles detected with  $|\theta_{\alpha}-\theta_{d}|=10^{\circ}$, for which there are no ambiguities in the number of counts and in the energy ($E_{d}$, e.g.) of each peak. 
A Monte Carlo simulation was performed in order to obtain the detection efficiency for $\alpha-d$ coincidences considering the experimental setup geometry.
All the necessary transformations from the laboratory frame to the appropriate center-of-mass frame were made following Refs.~\cite{Meijer85,Fuchs82} and also assuming an isotropic distribution of the breakup fragments in the $\alpha-d$ rest frame.
The corresponding CDCC results are represented by solid lines in Fig.~\ref{fig:BUdistrib}.

In Fig.~\ref{fig:BUdistrib} we can notice that the experimental angular distributions are well reproduced by CDCC calculations for each projectile energy.
The shape of a very similar angular distribution measured at $E_{lab}=41$~MeV by Bochkarev \textit{et al}.~\cite{Bochkarev85} is also well reproduced by CDCC~\cite{27} (see Fig. 3 of Ref.~\cite{27}). Thus, in order to calculate the $\sigma^{\rm exp}_{3^{+}}$ the CDCC angular distributions were normalized to the experimental ones. The values of the reduced chi-square ($\chi^{2}_{red}$) obtained from this procedure ($1.11\leq\chi^{2}_{red}\leq1.69$) are small enough to indicate a fair agreement between the experimental and the normalized theoretical results. The values of sequential breakup cross section obtained by integration of the normalized angular distributions ($\sigma^{\rm exp}_{3^{+}}$) and from the CDCC calculations ($\sigma^{\rm CDCC}_{3^{+}}$) for each energy are shown in Table~\ref{tab:NCBUexp}. When comparing our result at $E_{lab}=25.5$~MeV with the value of Ref.~\cite{17} for $^{6}$Li~+~$^{65}$Cu at $E_{lab}=25$~MeV ($22\pm2$~mb) we can notice that they are in agreement within the uncertainties.

In contrast to the $^{6}$Li~+~$^{65}$Cu~\cite{17} but in agreement with the $^{6}$Li~+~$^{28}$Si~\cite{18}, we have not observed in $^{6}$Li~+~$^{59}$Co any significant contribution of other $^{6}$Li resonant states (4.31~MeV 2$^{+}$ and 5.65~MeV 1$^{+}$) in our data. The comparison between the values of $\sigma^{\rm exp}_{3^{+}}$ or $\sigma^{\rm CDCC}_{3^{+}}$ and $\sigma_{NCBU}^{\rm CDCC}$ suggests a significant contribution of the direct breakup process in $\sigma_{NCBU}$, since the $\sigma_{NCBU}$ is the sum of the sequential and the direct breakup cross sections. This conclusion for the medium-mass target $^{59}$Co is rather consistent with either the stripping breakup mechanism proposed for the heavy $^{208}$Pb target~\cite{20} and/or with a competitive direct breakup for the light $^{28}$Si target~\cite{18}.

\section{Conclusions}

In this work we presented results for the intermediate mass target $^{6}$Li~+~$^{59}$Co reaction involving the weakly bound $^{6}$Li. Proton, deuteron and $\alpha$ particle inclusive measurements and $\alpha$-$d$ coincidence measurements were performed at the near barrier energies $E_{lab} = 17.4,~21.5,~25.5$ and $29.6~{\rm MeV}$. The contributions of different LCP production mechanisms were discussed. A statistical-model analysis, CDCC calculations and two-body kinematics were used as tools to provide information on the competing processes.

The analysis of the high-energy $\alpha$-bump and $d$-bump, obtained after the subtraction of the CF decay contribution, suggests that the main contribution to the high-energy $\alpha$-bump and $d$-bump cross sections is a combination of the ICF and TR mechanisms, as the non-capture BU cross section is estimated to be relatively small according to CDCC calculations. This assumption is confirmed firstly by the total $\alpha$ production, which is much more intense than predictions using the statistical model, and secondly by a two-body kinematics analysis. In this work it was not possible to fully disentangle the individual ICF and TR contributions. A clear separation of the different reaction mechanisms remains one of the main challenges in the study of fusion reactions induced by weakly bound nuclei. The difference between CDCC calculations for the non-capture breakup cross section and the experimental sequential cross section for the first excited $3^{+}$ ($E^{*}=2.186$~MeV) of $^{6}$Li (consistent with the sequential decay predicted by CDCC) suggests that the more significant contribution is due to the direct breakup process. However, as in the case of the $^{6}$Li~+~$^{28}$Si reaction~\cite{Pakou07} DWBA predictions of single nucleon TR cross section~\cite{27} are at least as large as direct BU cross sections.

\ack

The authors thank FAPESP and CNPq for financial support. N.K. gratefully acknowledges the receipt of a Marie Curie Intra-European Fellowship from the European Commission, contract No.\ MEIF-CT-2005-010158. We would also like to thank A. Pakou, A. M. Moro and C. Signorini for their careful reading of the manuscript.

\newpage 

\newpage 

\begin{table}
\caption{Summary of the results obtained from our analysis, showing for all the bombarding energies the total $\alpha$ and $d$ cross sections and the 
yields extracted from the $\alpha$-$bump$ and  $d$-$bump$, respectively. Experimental total fusion cross sections~\cite{Beck03}, total reaction cross 
sections from OM fits~\cite{31} and CDCC calculations~\cite{27} as well as the non-capture BU cross sections evaluated with CDCC calculations~\cite{27}
are also given.}
\begin{center}
\begin{tabular}{ c c c c c }
\hline\hline 
$E_{lab}$ (MeV) & $\sigma_{\alpha}^{total}$ (mb) & $\sigma_{d}^{total}$ (mb) & 
$\sigma_{\alpha-bump}$ (mb) & $\sigma_{d-bump}$ (mb) \\\hline 
17.4 & $404\pm22$ &  $86\pm8$  & $243\pm36$ &  $72\pm12$ \\
21.5 & $560\pm14$ & $140\pm10$ & $319\pm38$ & $107\pm13$ \\
25.5 & $715\pm29$ & $175\pm15$ & $332\pm33$ & $126\pm15$ \\
29.6 & $843\pm35$ & $217\pm15$ & $322\pm23$ & $150\pm18$ \\
\hline\hline
$E_{lab}$ (MeV) & $\sigma_{fus}^{\rm exp}$ (mb) & $\sigma_{Reac}^{\rm OM}$ (mb) & 
$\sigma_{Reac}^{\rm CDCC}$ (mb) & $\sigma_{NCBU}^{\rm CDCC}$ (mb) \\\hline 
17.4 &  $467\pm94$          &   780 &  943 & 33.6 \\
21.5 &     -                &  1099 & 1243 & 44.9 \\
25.5 &  $988\pm199$         &  1368 & 1430 & 54.7 \\
29.6 &      -               &  1540 & 1559 & 61.2 \\
\hline\hline
\end{tabular}
\end{center}
\label{tab:xsections}
\end{table}

\begin{table}
\caption{Experimental ($\sigma^{\rm exp}_{3^{+}}$) and CDCC calculations ($\sigma^{\rm CDCC}_{3^{+}}$) for the sequential breakup cross section for the first excited state $3^{+}$ ($E^{*}=2.186$~MeV) of $^{6}$Li. The $\sigma_{NCBU}^{\rm CDCC}$, already presented in Table~\ref{tab:xsections}, is shown again for comparison.}
\begin{center}
\begin{tabular}{ c c c c }
\hline\hline
$E_{lab}$ (MeV) & $\sigma^{\rm exp}_{3^{+}}$ (mb) & $\sigma^{\rm CDCC}_{3^{+}}$ (mb)& $\sigma_{NCBU}^{\rm CDCC}$ (mb) \\\hline
17.4 & $11.0\pm3.6$     & 17.1 & 33.6 \\
21.5 & $19.0\pm4.4$    & 21.0 & 44.9 \\
25.5 & $20.0\pm3.9$    & 22.9 & 54.7 \\
29.6 & $20.6\pm4.0$    & 23.5 & 61.2 \\
41.0 & $45\pm10^{a}$   & 22.5 & 79.4 \\
\hline\hline
\end{tabular}

\begin{footnotesize}$^{a}$ Experimental 2.18~MeV 3$^{+}$ sequential BU cross section reported in~\cite{Bochkarev85}.\end{footnotesize}
\end{center}
\label{tab:NCBUexp}
\end{table}

\end{document}